\documentclass[twocolumn]{article}

\usepackage{spconf}
\usepackage{url}
\usepackage{graphicx}
\usepackage{amsmath}
\usepackage{enumitem}

\begin{document}

\title{Improved multiple birdsong tracking with distribution derivative method and Markov renewal process clustering}

\author{}
\name{Dan Stowell$^{\star}$, Sa\v{s}o Mu\v{s}evi\v{c}$^{\dagger}$, Jordi Bonada$^{\dagger}$ and Mark D. Plumbley$^{\star}$}
\address{$^{\star}$Centre for Digital Music, Queen Mary University of London, UK \\
$^{\dagger}$Music Technology Group, Universitat Pompeu Fabra, Spain}

\maketitle

\begin{abstract}%
Segregating an audio mixture containing multiple simultaneous bird sounds is a challenging task.
However, birdsong often contains rapid pitch modulations, and these modulations carry information which may be of use in automatic recognition.
In this paper we demonstrate that an improved spectrogram representation, 
based on the distribution derivative method,
leads to improved performance of a segregation algorithm which uses a Markov renewal process model
to track vocalisation patterns consisting of singing and silences.
\end{abstract}

\section{Introduction}
\label{sec:intro}

Machine recognition of animal sounds is of growing importance in bioacoustics and ecology,
as a tool that can facilitate unattended monitoring, citizen science, and other applications with large volumes of audio data \cite{Eymann:2010,Walters:2012}.
For birdsong,
tasks which have been studied 
include recognition of species \cite{Stowell:2010e}
and individuals \cite{Fox:2008,Cheng:2012}.
However, much research considers only the monophonic case,
using recordings of single birds, either isolated or with low background interference.
It is important to develop techniques applicable to mixtures of singing birds: 
because singing often occurs within flocks or dawn choruses,
but also because there is research interest in analysing ensemble singing \cite{Malavasi:2012}
and in non-invasively characterising a 
population \cite{Fuller:2009}.
The automatic recognition literature has only just begun to approach such polyphonic tasks \cite{Briggs:2012}.

In the present work we focus on the task of analysing a recording containing multiple birds of the same species
(e.g.\ a recording of a flock),
and identifying the streams of \textit{syllables} that correspond to a single bird.
From the perspective of \textit{computational auditory scene analysis}
this task of clustering sounds is
analogous to the well-known ``cocktail party problem'' in perception \cite{Wang:2006}.
We consider the task recently studied by \cite{Stowell:2013},
which develops a probabilistic model that can segregate such sequences of sound events
modelled as point processes.
In that work, it was observed that the quality of the initial detection stage (used to locate individual syllables)
when applied to audio mixtures can be a strong limiting factor on the quality of the tracking.
In this paper we work within the same paradigm 
and demonstrate that improvements to the underlying representation yield improved quality of tracking.

In \cite{Stowell:2012c} it was observed that birdsong contains
very rapid modulations, and that using a chirplet representation instead of standard spectral magnitudes
could lead to improved recognition performance by making use of low-level modulation information.
The technique described in that paper used a simple dictionary of chirplets to analyse a signal.
However, powerful parametric techniques exist to estimate the characteristics of non-stationary signals and may be well-suited to this task.
The generalised reassignment method (GRM) \cite{Wen:2009} has be shown to work well for this even when dealing with extreme frequency and amplitude modulations 
\cite{Musevic:2011}. However difficulties arise as the linear system of equations for a third degree GRM becomes ill-conditioned.
A related method, the distribution derivative method (DDM) \cite{Betser:2009} circumvents this. 
In addition a frequency range, rather than just a single frequency 
can be examined when a highly modulated sinusoid is assumed to occupy a significant portion of spectrum, rather than being concentrated around the peak frequency. 

Such techniques have not yet been widely tested in practical applications.
In the present work we demonstrate that the refined representation derived from the DDM leads to improved tracking of multiple singing birds.
In the remainder of this paper, we will give an overview of the DDM, and the particular variant of the technique developed for the present study.
We will then describe the multiple tracking technique used to infer the sequence structure contained within a recording of multiple birds.
We will apply this tracking procedure to a dataset of birdsong recordings,
analysed via either a standard spectrogram or the DDM,
showing that the improved spectral representation is of benefit to downstream analysis.

\section{Distribution derivative method}
\label{sec:ddm}
The essence of the DDM lies in a simple but powerful concept of the distribution derivative rule. 
Considering an arbitrary distribution $x$ and a test function $\Psi$,  a straightforward consequence using \textit{integration-per-partes} on \textit{inner product} follows:
\begin{equation}
<x',\Psi> = -<x,\Psi'>.
\end{equation}
Treating the signal under study as a distribution, the following equality can be obtained using the above:
\begin{equation}
<s',w e^{j\omega}>=-<s,w'e^{j\omega}> + j \omega <s,w e^{j\omega}>,
\end{equation}
where $<.,.>$ denotes the inner product, $w$ the window function of finite time support and $s$ the signal. In such a setting the Fourier Transform (FT) at frequency $\omega$ can be written as:
\begin{equation}
\label{eq:ft_inner}
S_{w}(\omega) = <s,w e^{j\omega}>.
\end{equation}
If the signal is modelled as a \textit{generalised sinusoid}:
\begin{equation}
s(t) = e^{r(t)},r(t) = \sum_{k=0} r_{k} t^{k}, r_{k} \in \mathbf{C} ,
\end{equation}
the following equality:
\begin{align}
S'_{w}(\omega) =& <r' s,w e^{j\omega}> \\ 
                         =& \sum_{k=0}^{K-1} (k+1) r_{k+1} <t^{k} s,w e^{j\omega}> \\                         
                         =& -<s,w' e^{j\omega}> + j \omega <s,w e^{j\omega}>,
\end{align}
can be compacted (using \eqref{eq:ft_inner}) into:
\begin{equation}
\sum_{k=0}^{K-1} (k+1) r_{k+1} S_{t w}(\omega)  = -S_{w'}(\omega) + j \omega S_{w}(\omega),
\end{equation}
where $<s t^{k}, w e^{j\omega}> = <s, (t^{k} w) e^{j\omega}>=S_{t^{k} w}$.
The above holds for any $\omega$ and can thus be used to define a linear system of equations with respect to $r_{k},k>0$; however $r_{0}$ cannot be estimated this way as it was factored out during derivation. The frequencies used to construct the linear system can generally be arbitrary, though one should choose the ones that bear most of the energy of the sinusoids under study to avoid numerical instabilities. The set of frequencies should also cover a big part of the bandwidth occupied by the sinusoid: failure to do so would exclude important frequency domain content of the sinusoid, leading to inaccurate estimation.

 The complex stationary parameter $r_{0}$ can be estimated after the non-stationary parameters $r_{k},k>0$ have been estimated. Substituting $<s,s>=< e^{ \Re (r(t))}, e^{\Re(r(t))}>$ into:
\begin{equation} \label{eq:amp_dmd_krnl_a}
<s,e^{\sum_{k=1}r_{k}t^{k}}> =e^{r_{0}} < e^{ \Re (r(t))}, e^{\Re(r(t))}>,
\end{equation}
yields:
\begin{equation} \label{eq:amp_dmd_krnl}
e^{r_{0}} = \frac{<s,e^{\sum_{k=1}r_{k}t^{k}}>}{<s,s>}.
\end{equation}
The parameters $r_{k},k>0$ in \eqref{eq:amp_dmd_krnl} are substituted with estimates $\hat{r}_{k},k>0$ to get the estimate for $e^{r_{0}}$. For most applications the model with FM polynomial of degree 2 (i.e.\ frequency change is linear during the observation frame) is sufficient. In such case only values $S_{w},S_{w'},S_{tw}$ at different frequencies form the linear system.The widest mainlobe width is that of $w(t) t$, 5 bins in total.
In order to select an optimal bandwidth for the DDM, a typical bird frequency change must be estimated from a real recording. 
For the Chiffchaff sounds considered in this paper, a typical chirp exhibits a maximum of 100 kHz/s frequency change, thus for an observation frame of 1024 samples and a sampling frequency of 44100 such chirp would cover roughly about 2300Hz bandwidth. It will be shown that covering a region of about 1000Hz is sufficient to estimate the linear FM accurately enough for the purpose.

To use DDM efficiently we use the bins of the FFT: computing DFT at an arbitrary frequency has little advantage and increases computational load significantly. For this paper a bandwidth of 16 bins was considered. The cumulative effective range of all mainlobes positioned at 16 consecutive bins therefore totals to 21 bins (since the widest mainlobe width is 5 bins), almost 1000Hz in the current setting. 

\begin{figure}
	\includegraphics[width=7.9cm, clip, trim=12mm 10mm 26mm 8mm]{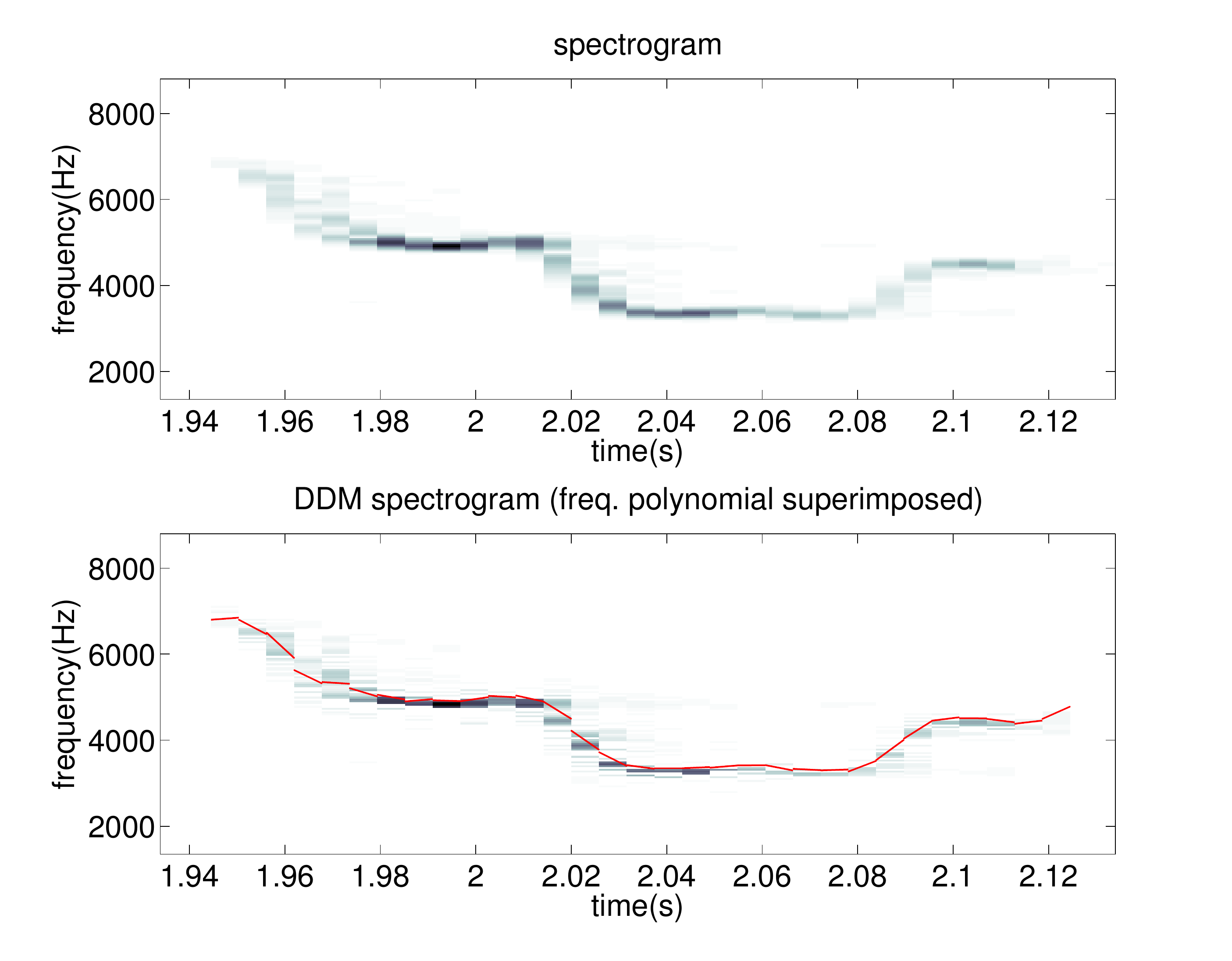}
\caption{Top: spectrogram, Bottom: DDM spectrogram with linear freq polynomial of magnitude peak superimposed.} 
\label{figspecr_reass_poly}
\end{figure}

The estimates depend on the frequency range examined, so we designate them as $\hat{r}_{k}(\omega_{L},\omega_{H})$. The frequency and amplitude estimates $\Im(\hat{r}_{1}(\omega_{L},\omega_{H})), \Re(\hat{r}_{0}(\omega_{L},\omega_{H}))$ are of particular interest as a time-frequency representation (TFR) based on \textit{reassignment} \cite{Auger:1995} can be constructed. The frequency estimate is generally not an exact bin value: quantising and summing the corresponding amplitude estimates results in a TFR very similar to the \textit{reassigned spectrogram} \cite{Auger:1995}, which we call the \textit{DDM spectrogram}.
It will be shown that such TFR exhibits desirable properties
, especially when combined with the linear frequency change estimate $\Im(\hat{r}_{2})$.

\section{Multiple tracking with a Markov renewal process model}
\label{sec:mmrp}

The task of tracking multiple sound sources in an acoustic scene
may be approached using established multiple-tracking paradigms
\cite{Mahler:2007a}.
However, such models do not account for structured patterns of
emission and silence, as is common in many sound types including birdsong.
For example the factorial hidden Markov model
does not formally model gaps
although silent states can be added to the representation \cite{Mysore:2012};
however it
assumes an unchanging number of sources.

In order to track a varying number of intermittent sources,
\cite{Stowell:2013}
introduced a multiple tracking model with sources modelled as
instances of a \textit{Markov renewal process}.
A Markov renewal process is a point process
in which the current state stochastically determines the following state,
as well as the time gap between them:
\begin{align}
      P(\tau_{n+1} \le t, X_{n+1}=j|(X_1, T_1),\ldots, (X_n=i, T_n)) & \nonumber \\
   =  P(\tau_{n+1} \le t, X_{n+1}=j|X_n=i) & \nonumber \\
                   \qquad \qquad
                   \, \forall n \ge1, \,  t\ge0, \,   i,j \in \mathcal{S} ,
\label{eq:mrp}
\end{align}
where 
observations are received in the form 
$\{(X, T)\}$ with state $X$ and time $T$,
and 
$\tau_{n+1}$ is the time difference $T_{n+1} - T_n$.

Note that $\tau$ is known if the observations represent a single sequence,
but if the observations may represent multiple sequences as well as clutter noise
then the causal structure is unknown and $\tau$ is hidden.
In that case we can estimate the structure by 
choosing a partitioning of the data into $K$ clusters plus $H$ noise events
so as to 
maximise the likelihood
\begin{align}
        \textrm{L} = %
        \prod_{   k=1}^K{p_{\textrm{MRP}}(k)}  %
        \prod_{\eta=1}^H{p_{\textrm{NOISE}}(\eta)}  , \nonumber
\end{align}
where
$p_{\textrm{MRP}}(k)$
represents the likelihood of the observation subsequence in cluster $k$ being generated by a single MRP, 
with internal transition likelihoods as in \eqref{eq:mrp}, 
and
$p_{\textrm{NOISE}}(\eta)$
represents the likelihood of a single noise datum.
In this multiple Markov renewal process (MMRP) setting,
inferring the maximum likelihood solution is a combinatorial problem
which can be addressed via graph-theoretic techniques \cite{Stowell:2013}.

The MMRP inference technique does not operate directly on audio,
but takes a set of timestamped event detections as input.
In \cite{Stowell:2013} the authors describe an experiment applied to birdsong,
in which they use a simple cross-correlation signal detection technique
applied to spectrogram data as the preprocessing step for their analysis.
They observe that this step may be a limiting factor in overall performance,
in part because the cross-correlation may not recover the same detections 
from audio mixtures as from monophonic audio,
and in part because each detection has only a simple state representation
(frequency offset of the template match).

Various refinements might be tried to improve on the results of \cite{Stowell:2013},
such as alternative event detection based on dictionary learning techniques
or sinusoidal modelling.
However, in this work we will use the same detection technique as the previous authors,
and demonstrate that using the distribution derivative method (DDM) of Section \ref{sec:ddm}
improves the recovery of birdsong sequences within the same workflow,
by improving the spectrotemporal detail in the underlying representation.

\section{Experiments}
\label{sec:expt}

\providecommand{\fsn}{F_\text{SN}}
\providecommand{\ftrans}{F_\text{trans}}

To validate the MMRP inference, the authors in \cite{Stowell:2013}
apply it to a dataset of birdsong audio files,
using the solo files as training data and mixed audio files to test 
segregation of bird sounds into separate streams.
We ran the same experiment, varying 
the underlying spectrogram representation and the amount of detail
passed on to the later processing stages.


\begin{figure*}[t!]  
	\includegraphics[width=0.288\textwidth, clip, trim=10mm 17mm 20mm 15mm]{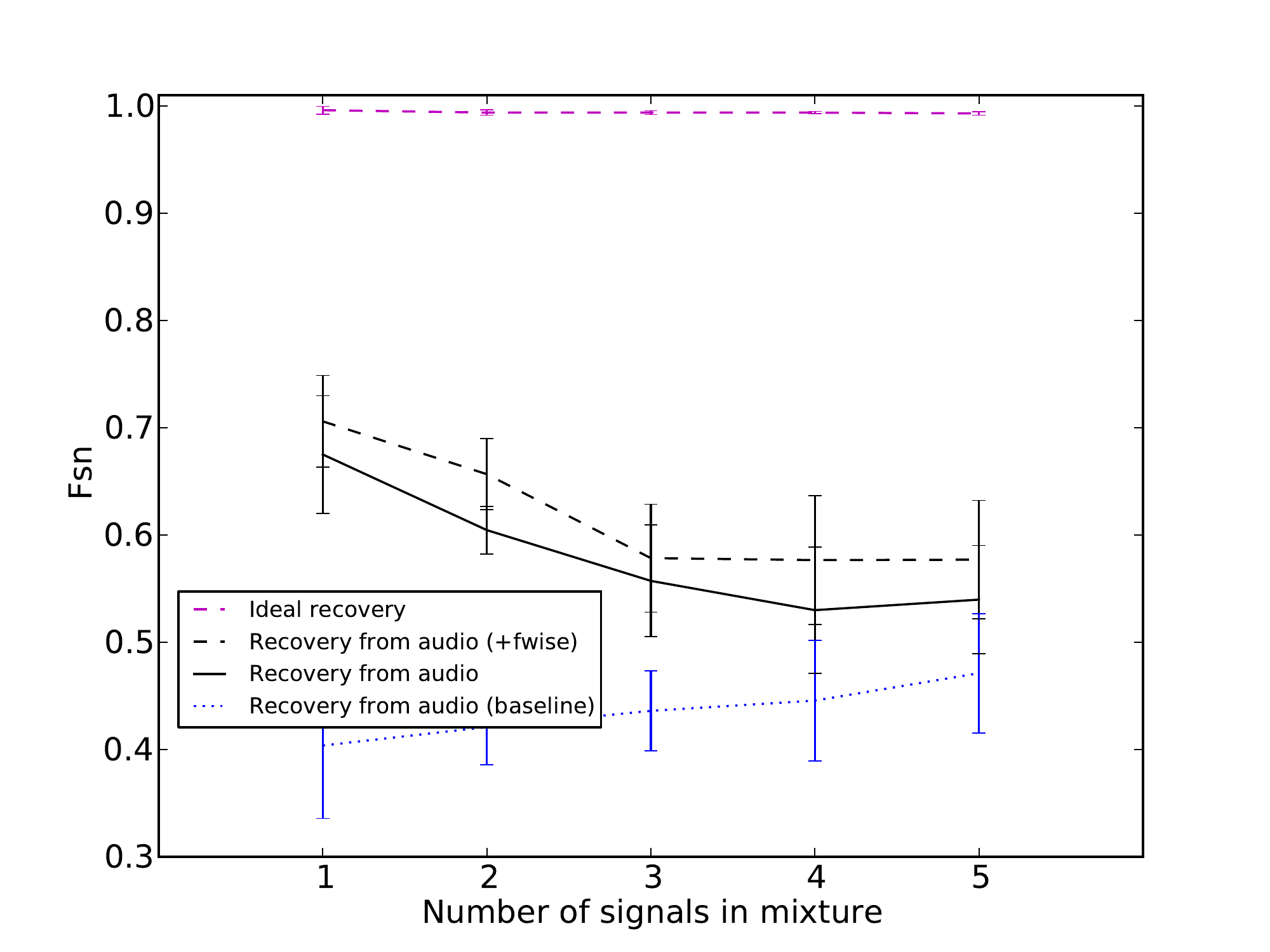}%
	\includegraphics[width=0.261\textwidth, clip, trim=26mm 17mm 20mm 15mm]{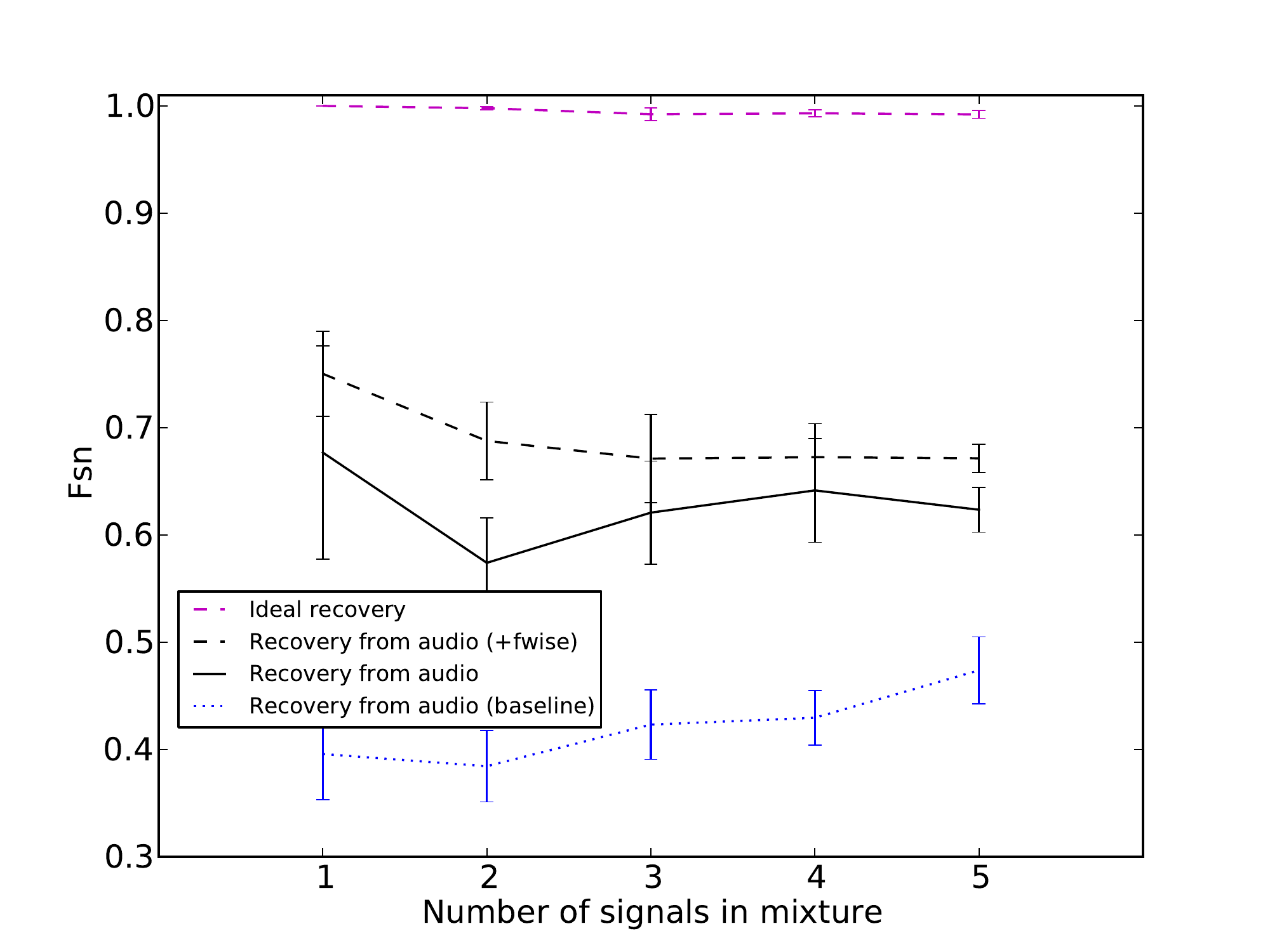}%
	\includegraphics[width=0.261\textwidth, clip, trim=26mm 17mm 20mm 15mm]{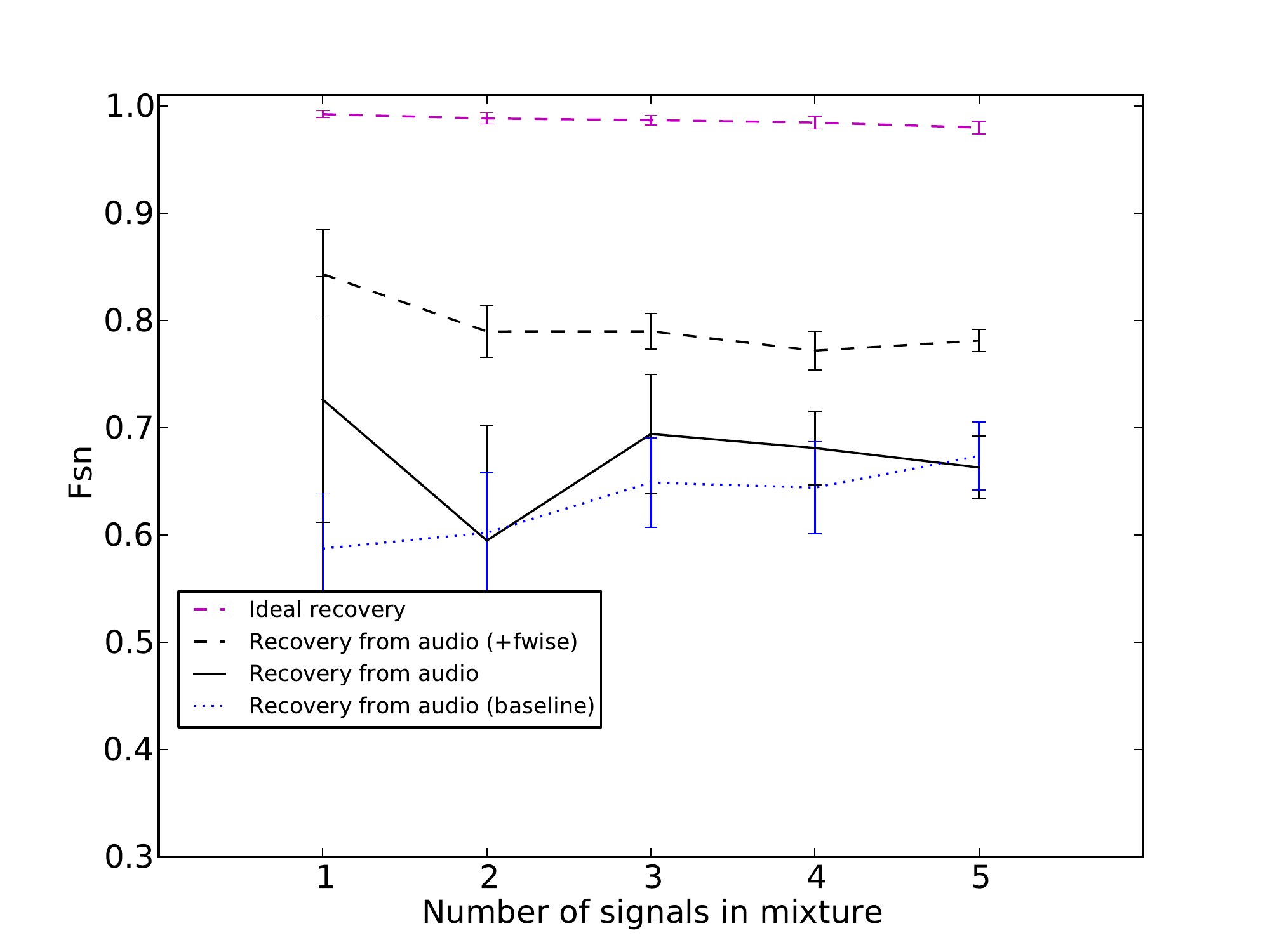}%
	\\%
	\includegraphics[width=0.288\textwidth, clip, trim=10mm 3mm 20mm 15mm]{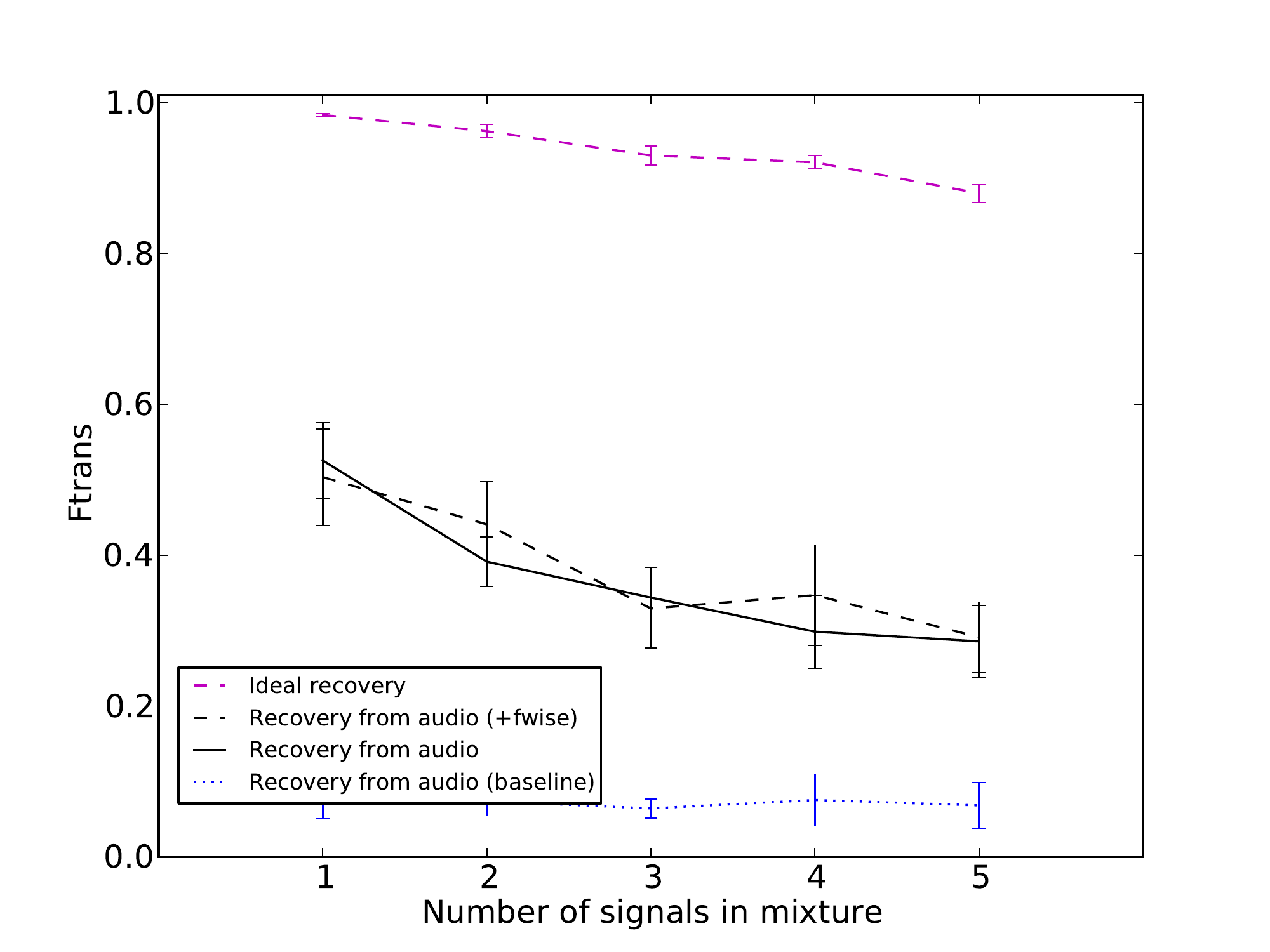}%
	\includegraphics[width=0.261\textwidth, clip, trim=26mm 3mm 20mm 15mm]{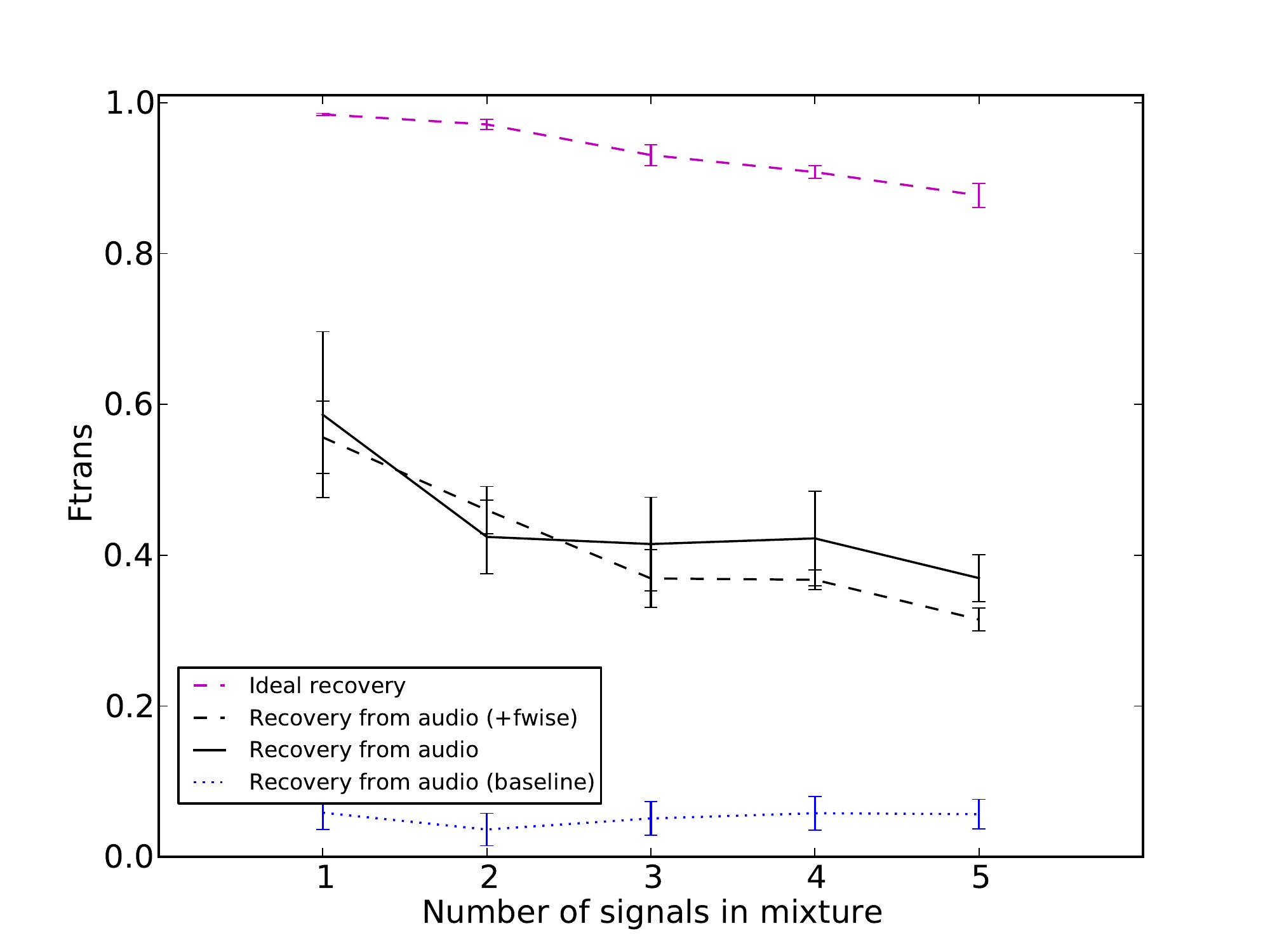}%
	\includegraphics[width=0.261\textwidth, clip, trim=26mm 3mm 20mm 15mm]{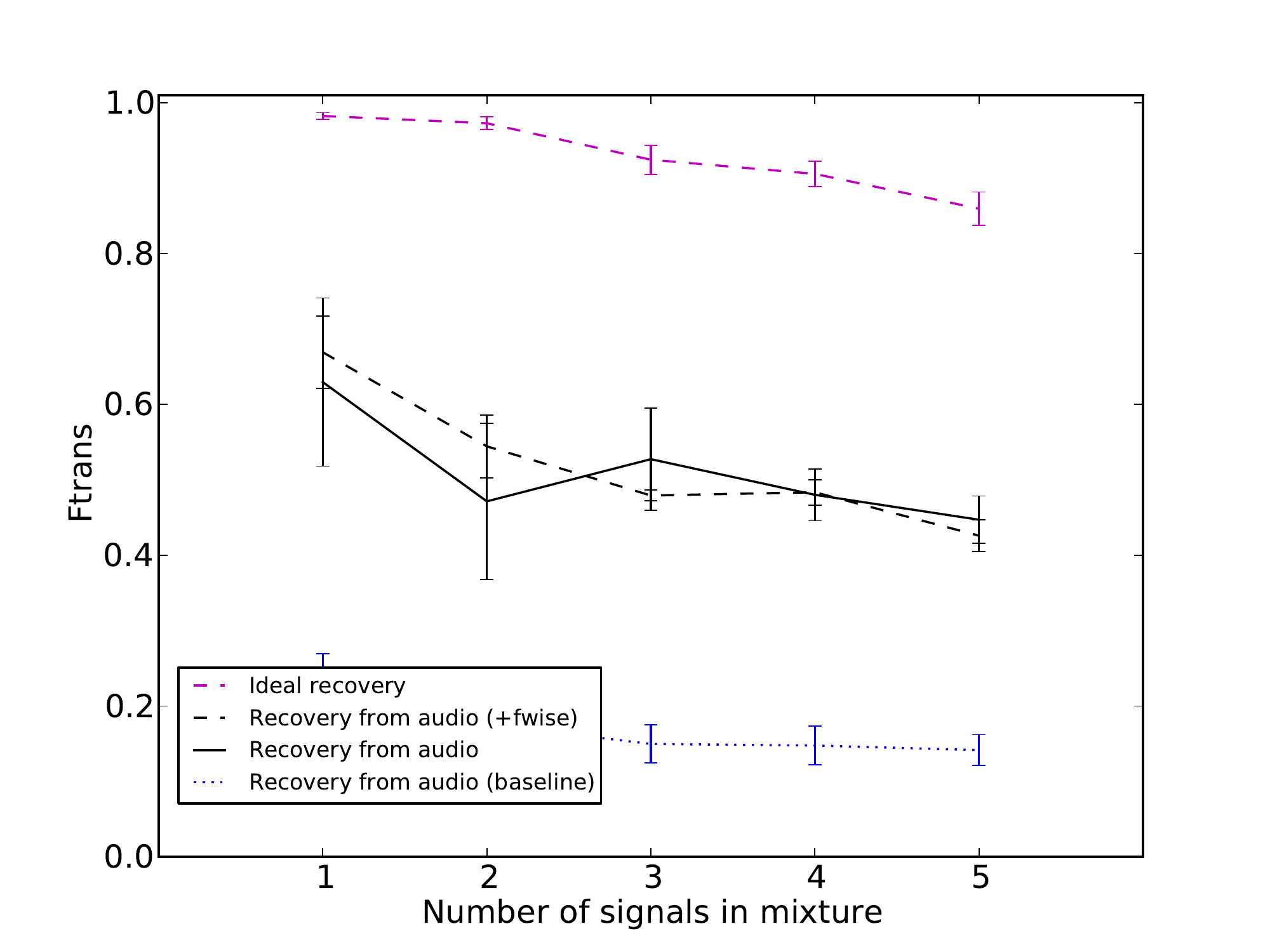}%
	\caption{%
F-measure statistics for signal-noise separation ($\fsn$, top row) and recovery of transitions ($\ftrans$, bottom row). %
The three columns show results using the three different signal representations: standard STFT spectrogram (left), DDM (middle), and DMM including first-order FM information (right). %
The solid black line shows performance using the standard encoding of each detection as a single value, %
while the dashed black line shows performance using the more detailed encoding with five frequency values per syllable. %
Means and standard errors are shown, five-fold crossvalidation. %
}
	\label{fig:fsnftrans}
\end{figure*}

The dataset consists of individual recordings of the Common Chiffchaff
(\textit{Phylloscopus collybita})
collected around Europe and submitted to the public database
Xeno Canto.%
\footnote{\url{http://www.xeno-canto.org/europe}} %
The specific recordings used are available online.%
\footnote{\url{http://archive.org/details/chiffchaff25}} %
For each recording, as well as for the synthetic mixtures,
we analysed the audio via a standard STFT spectrogram
(sample rate 44.1 kHz, frame size 1024, 50\% overlap, Hann window), 
and separately via the DDM spectrogram described in Section \ref{sec:ddm}.

We used the same cross-correlation template-matching paradigm as \cite{Stowell:2013}
to detect individual syllables of birdsong.
To detect syllables from the standard spectrogram we used the same manually-specified template.
Additionally we tested two strategies to detect syllables from the DDM spectrogram:
the standard 2D time-frequency template,
or a 3D time-frequency-FM template created by augmenting the template
with a third dimension representing the FM values expected in the syllable.
These FM values were calculated from the frequency slope implied by the shape of the template.

By testing these three variants of the template-matching process 
(STFT, DDM, DDM with FM information),
we could evaluate whether improving the detection could have positive effects on the birdsong segregation.
However, we also wanted to investigate whether the MMRP segregation process
would be improved by giving it access to a more detailed representation
of each detected syllable.
To that end, we tested the approach of \cite{Stowell:2013},
which encodes each syllable state $X$ simply as a single freqeuency-offset value,
against a modified approach in which spectral detail from within the detection region is encoded as a vector-valued state.
For each detected syllable, we determined a simple feature representing the time evolution of 
the spectral energy in the detection region:
the frequency of the peak bin in each frame,
downsampled by a factor of four to alleviate curse-of-dimensionality concerns.
This produced a vector of five frequency values for each detection.
Our hypothesis was that this richer representation would allow the 
MMRP inference to make clearer distinctions between true and false transitions,
and thus improve the performance.

In our tests we used the same baseline and gold-standard systems as in \cite{Stowell:2013}.
The baseline was based on a Gaussian mixture model (GMM) signal-vs-noise classifier which
does not make use of transition likelihoods;
the gold standard was not to use the mixture audio files as input,
but to combine the detections produced from the monophonic file analysis,
representing the event detections that would be recovered in the ideal case
that detections from an audio mixture are the same as those from the separate audio signals.
However, our main point of comparison was between the different underlying representations,
to examine whether the improved spectrogram and/or the more detailed output improves performance.

As in \cite{Stowell:2013}, we performed five-fold crossvalidation,
with the standard F-measure as our evaluation statistic
applied in two ways:
$\fsn$ is the F-measure for signal/noise separation,
and 
$\ftrans$ is the F-measure for recovering true event-to-event transitions
(i.e.\ segregating the signal correctly into sources).

Results are shown in Figure \ref{fig:fsnftrans}.
It is evident from the graphs that performance improves from the left plots to the right plots:
using DDM rather than the STFT spectrogram improves performance,
and using DDM with the FM information included in the detection step improves it further still.
This applies for both $\fsn$ and $\ftrans$.
(Interestingly, the use of DDM with FM information also improves the performance of the baseline non-MMRP inference.)
However, the effect of passing the more detailed state representation in to the MMRP inference
(the solid lines vs.\ the dashed lines)
appears to improve $\fsn$ without notably changing $\ftrans$.

We confirmed these observations using a repeated-measures ANOVA test.
For each evaluation measure we entered three factors: the spectrogram type, the state representation, and the number of signals in the mixture.
For $\fsn$, significant effects were found for all three factors (each significant at $p<0.006$).
For $\ftrans$, significant effects were found for the spectrogram type and the number of signals in the mixture (each $p<0.007$), but the state representation was not significant ($p=0.056$).
For both evaluation measures, a significant two-way interaction was also found for spectrogram mode combined with number of signals ($p<0.007$).

Overall, in this experiment we achieved around 20 percentage point improvements in both $\fsn$ and $\ftrans$, 
using a combination of the DDM spectrogram, the use of FM information in template-matching, and passing a more detailed state representation to the source-segregation stage.

\section{Conclusions}
\label{sec:conc}

We have considered a maximum-likelihood technique for tracking multiple singing birds in an audio recording,
and demonstrated that it can benefit strongly from an improved underlying spectrogram representation.
We applied a variant of the DDM technique, using a range of spectral bins to infer 
fine detail about modulated sinusoids,
which is particularly pertinent in the case of birdsong because of the presence of rapid pitch modulations.
We also demonstrated that passing a rich feature representation to the later inference stage
also improves tracking.
Altogether, our modifications yield approximately 20 percentage point improvement in the F-measure.


\section{Acknowledgments}

DS \& MP are supported by an EPSRC Leadership Fellowship EP/G007144/1.

\clearpage
\bibliographystyle{IEEEbib}
\bibliography{bibliog,bibliog_ddm}

\begin{thebibliography}{10}

\bibitem{Eymann:2010}
J.~Eymann, J.~Degreef, C.~H{\"a}user, J.~C. Monje, Y.~Samyn, and D.~Van~den
  Spiegel,
\newblock {\em Manual on field recording techniques and protocols for All Taxa
  Biodiversity Inventories},
\newblock ABC Taxa. Belgian National Focal Point for the GTI, Brussels, 2010.

\bibitem{Walters:2012}
C.L. Walters, R.~Freeman, A.~Collen, C.~Dietz, M.~Brock~Fenton, G.~Jones, M.K.
  Obrist, S.J. Puechmaille, T.~Sattler, B.M. Siemers, et~al.,
\newblock ``A continental-scale tool for acoustic identification of european
  bats,''
\newblock {\em Journal of Applied Ecology}, vol. 49, pp. 1064--1074, 2012.

\bibitem{Stowell:2010e}
D.~Stowell and M.~D. Plumbley,
\newblock ``Birdsong and {C4DM}: A survey of {UK} birdsong and machine
  recognition for music researchers,''
\newblock Tech. {R}ep. C4DM-TR-09-12, Centre for Digital Music, Queen Mary
  University of London, Aug 2010.

\bibitem{Fox:2008}
E.~J.~S. Fox,
\newblock {\em Call-independent identification in birds},
\newblock Ph.D. thesis, University of Western Australia, 2008.

\bibitem{Cheng:2012}
J.~Cheng, B.~Xie, C.~Lin, and L.~Ji,
\newblock ``A comparative study in birds: call-type-independent species and
  individual recognition using four machine-learning methods and two acoustic
  features,''
\newblock {\em Bioacoustics}, vol. 21, no. 2, pp. 157--171, 2012.

\bibitem{Malavasi:2012}
R.~Malavasi and A.~Farina,
\newblock ``Neighbours' talk: interspecific choruses among songbirds,''
\newblock {\em Bioacoustics}, 2012.

\bibitem{Fuller:2009}
R.A. Fuller, J.~Tratalos, and K.J. Gaston,
\newblock ``How many birds are there in a city of half a million people?,''
\newblock {\em Diversity and Distributions}, vol. 15, no. 2, pp. 328--337,
  2009.

\bibitem{Briggs:2012}
F.~Briggs, B.~Lakshminarayanan, L.~Neal, X.Z. Fern, R.~Raich, S.J.K. Hadley,
  A.S. Hadley, and M.G. Betts,
\newblock ``Acoustic classification of multiple simultaneous bird species: A
  multi-instance multi-label approach,''
\newblock {\em The Journal of the Acoustical Society of America}, vol. 131, pp.
  4640--4650, 2012.

\bibitem{Wang:2006}
D.~L. Wang and G.~J. Brown, Eds.,
\newblock {\em Computational Auditory Scene Analysis: Principles, Algorithms,
  and Applications},
\newblock Chapter 1. IEEE Press, 2006.

\bibitem{Stowell:2013}
D.~Stowell and M.~D. Plumbley,
\newblock ``Segregating event streams and noise with a {M}arkov renewal process
  model,''
\newblock submitted,
\newblock preprint arXiv:1211.2972.

\bibitem{Stowell:2012c}
D.~Stowell and M.~D. Plumbley,
\newblock ``Framewise heterodyne chirp analysis of birdsong,''
\newblock in {\em Proceedings of EUSPICO}, 2012.

\bibitem{Wen:2009}
X.~Wen and M.~Sandler,
\newblock ``Notes on model-based non-stationary sinusoid estimation methods
  using derivative,''
\newblock in {\em Proceedings of the 12th Int. Conference on Digital Audio
  Effects (DAFx-09)}, 2009.

\bibitem{Musevic:2011}
S.~Mu{\v{s}}evi\v{c} and J.~Bonada,
\newblock ``Generalized reassignment with an adaptive polynomial-phase
  {F}ourier kernel for the estimation of non-stationary sinusoidal
  parameters,''
\newblock {\em Proc. Digital Audio Effects (DAFx), Paris, France}, pp.
  317--374, 2011.

\bibitem{Betser:2009}
M.~Betser,
\newblock ``Sinusoidal polynomial parameter estimation using the distribution
  derivative,''
\newblock {\em {IEEE} Transactions on Signal Processing}, vol. 57, no. 12, pp.
  4633 --4645, dec. 2009.

\bibitem{Auger:1995}
F.~Auger and P.~Flandrin,
\newblock ``Improving the readability of time-frequency and time-scale
  representations by the reassignment method,''
\newblock {\em {IEEE} Transactions on Signal Processing}, vol. 43, no. 5, pp.
  1068--1089, 1995.

\bibitem{Mahler:2007a}
R.~P.~S. Mahler,
\newblock {\em Statistical Multisource-Multitarget Information Fusion},
\newblock Artech House, Boston/London, 2007.

\bibitem{Mysore:2012}
G.~Mysore and M.~Sahani,
\newblock ``Variational inference in non-negative factorial hidden {M}arkov
  models for efficient audio source separation,''
\newblock in {\em Proceedings of the International Conference on Machine
  Learning (ICML)}, Edinburgh, Scotland, June 2012.

\end{thebibliography}
\end{document}